\documentclass{article}
\usepackage[T1]{fontenc} 
\usepackage[utf8]{inputenc} 
\usepackage{ismir,amsmath,cite,url}
\usepackage{amssymb}
\usepackage{stmaryrd}
\usepackage{graphicx}
\usepackage{subcaption}
\usepackage{xcolor}
\usepackage[separate-uncertainty=true]{siunitx}
\usepackage{array, booktabs}
\usepackage{makecell}
\usepackage{rotating}
\usepackage{multirow}
\usepackage{tabularray}

\newcommand{\nobend}{$\varnothing$~}
\newcommand{\up}{$\uparrow$~}
\newcommand{\held}{$\rightarrow$~}
\newcommand{\down}{$\downarrow$~}

\title{Modeling bends in popular music guitar tablatures}

\oneauthor
  {Alexandre D'Hooge \hspace{1cm} Louis Bigo \hspace{1cm} Ken Déguernel}
  {Univ. Lille, CNRS, Centrale Lille, UMR 9189 CRIStAL, F-59000 Lille, France\\
  {\tt alexandre.dhooge@algomus.fr}}

\def\authorname{A. D'Hooge, L. Bigo and K. Déguernel}

\usepackage[bookmarks=false,pdfauthor={\authorname},pdfsubject={\papersubject},hidelinks]{hyperref}

\begin{document}

\maketitle
\begin{abstract}
Tablature notation is widely used in popular music 
to transcribe and share guitar musical content. 
As a complement to standard score notation, 
tablatures transcribe performance gesture information 
including finger positions and a variety of 
guitar-specific playing techniques such as 
\emph{slides}, \emph{hammer-on/pull-off} or \emph{bends}.
This paper focuses on bends,  
which enable to progressively shift the pitch of a note, 
therefore circumventing physical limitations 
of the discrete fretted fingerboard. 
In this paper, we propose a set of 25 high-level features, 
computed for each note of the tablature, 
to study how bend occurrences can be predicted 
from their past and future short-term context. 
Experiments are performed on a corpus of 
932 \emph{lead guitar} tablatures of popular music 
and show that a decision tree successfully predicts 
bend occurrences with an F$_1$ score of \num{0.71} and
a limited amount of false positive predictions, 
demonstrating promising applications to assist 
the arrangement of non-guitar music into guitar tablatures.

\end{abstract}

\section{Introduction}\label{sec:introduction}

The guitar, whether acoustic or electric, is  
(in most cases) a fretted instrument which enforces 
the playing of discrete pitch values on a chromatic scale. 
This constraint can be beneficial, as it limits the risk of playing out-of-tune. 
However, it also prevents microtonal experiments 
or continuous pitch shifts, which can be 
a powerful means of musical expressiveness. 
To overcome this limitation, guitarists can alter 
the string tension with their fretting hand \cite{grimes_string_2014}
to reach a completely new pitch, up to several semitones higher. 
This technique is called string bending, or just \emph{bends,} 
and is an important part of guitar playing in blues, rock or pop music.
Even though bending a string can only increase the pitch of a note, 
a variety of bend types are used.
While guitar players mostly agree over the existing variations, the names used can differ. 
In this paper, we consider the five bend types 
described in Gomez's work \cite{gomez_modern_2016}:
\emph{Basic upward}, \emph{Held},
\emph{Reverse}, \emph{Up \& Down}, and
\emph{Complex} bends for bends that do not belong to any of the previous categories.

\begin{figure*}
\centering
	\includegraphics[width=\textwidth]{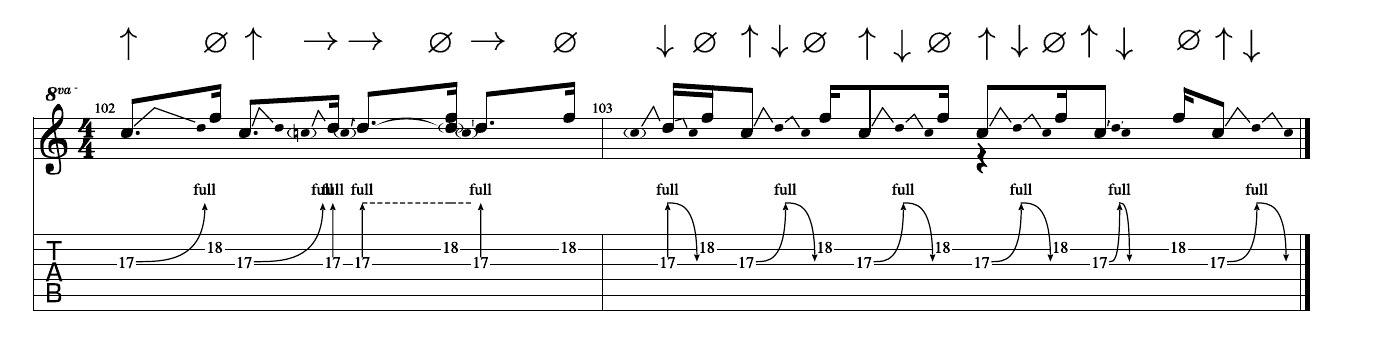}
	\caption{Excerpt from Lynyrd Skynyrd's \textit{Free Bird} solo. In the first measure,
 the first two bends are \emph{basic upward bends}, the remaining ones are \emph{held bends}. 
 In the second measure, the first bend is a \emph{reverse bend}, and the other ones are \emph{up and down bends}. While all bends of this example are whole tones (denoted by \emph{full}), the amplitude of a bend can vary. String movements are labeled above, according to the representation presented in \autoref{sec:model}. An audio rendering of this excerpt is available on the \href{https://gitlab.com/adhooge1/bend-prediction/-/blob/main/media/free-bird.mp4}{\textbf{accompanying repository}}.
 }
	\label{fig:bendTypes}
\end{figure*}

Guitar tablatures, compared to standard staff notation, include fingering information
on where to play a note with a given pitch on the fretboard.
Tablatures are therefore an effective notation to display playing techniques,
the position of the fretting hand being critical to know how to perform a bend.
 Examples of bends are shown in a tablature in \autoref{fig:bendTypes},
different bend types are represented with differently shaped arrows. 
Knowing where and when to use bends is part of the idioms of guitar pop music
and an important part of learning this style. However, the variety of bend types
can make it difficult to choose how to use them. For instance,
a guitarist playing a score that was composed for another instrument
may want to add expressiveness using bends, and could need help on deciding
which notes to bend and which bend to use. Moreover, a tool suggesting bends could
also improve the quality of online tablatures that sometimes do not have
guitar techniques annotations.

Could bends be inferred from musical context? Are they correlated to other 
elements of a score or a tablature such as pitch, rhythm, or hand position? 
In this paper, we propose to model bent notes and their context
through temporal, pitch and tablature related information. Such a representation could be used
to predict which notes are bent from a score or a tablature.
Our contribution is three-fold: (1) we define a set of high-level
features to model bent notes and their context, 
(2) we conduct a statistical study on bends based on those features, and
(3) we propose a method for predicting bends from tablatures.

The rest of this paper is organized as follows: 
after discussing related works in \autoref{sec:SOTA}, we introduce our modeling choices in
\autoref{sec:model}. The dataset and its statistical study are presented in \autoref{sec:dataset}.
We introduce our prediction algorithm in \autoref{sec:results} and reflect on this work in \autoref{sec:discussion}.

%
\section{Related Works}
\label{sec:SOTA}

Guitar tablatures are often studied in the audio realm
for guitar music transcription \cite{xi_guitarset_2018, wiggins_guitar_2019}.
In particular, automatic transcription of playing techniques from audio has also been studied,
for instance with hexaphonic microphones \cite{reboursiere_left_2012} 
or deep learning techniques \cite{chen_playing_2018}.
Of course, this task is not restricted to guitar 
and has been applied to other instruments 
such as the Chinese \textit{guqin}, which also features 
several string bending techniques \cite{yu-fen_huang_joint_2020}. 

Another related task is the generation of tablatures, which has been studied 
increasingly in the last decade. Playing techniques can be included in generation 
frameworks to obtain results closer to actual human performance. The Transformer-XL model
presented in \cite{sarmento_dadagp_2021, sarmento_gtr-ctrl_2023} can for instance generate tokens representing playing techniques.
Chen et al. \cite{chen_automatic_2020} also consider \textit{Technique} events, but only
for picking-hand techniques, since the generation focuses on fingerstyle guitar.
McVicar et al.  present in \cite{mcvicar_autoleadguitar_2014} a method to generate tablatures of guitar solo phrases, and fretting-hand techniques are added in a post-processing step. Beyond the case of guitar, playing techniques modeling includes symbolic piano music generation with sustain pedal information \cite{ching_learning_2021}.

Another large part of guitar tablature MIR research focuses on fingering prediction, 
where a model is designed to predict the pitch/fret combination 
for each note of a musical score. 
This problem has been studied with path-finding algorithms \cite{sayegh_fingering_1989} 
and minimax techniques \cite{hori_minimax_2016}.
More recently, Cheung et al. \cite{cheung_semi-supervised_2021} used a deep learning model 
to generate fingering annotations, though not for guitar but violin.
Those instruments nonetheless share some properties, and a study of fingering prediction on string instruments like the violin or the guitar,
compared to other instruments, can be found in \cite{hori_three-level_2021}. 
However, the task of using symbolic music to study occurrences of playing techniques is rarely studied. 
Xie and Li \cite{xie_symbolic_2020} propose to predict playing techniques as a 
tagging task on symbolic bamboo flute music but, to the best of our knowledge, 
no such work has been proposed for guitar music.

\section{Modeling bends in tablature}
\label{sec:model}

To formulate bend prediction as a machine learning task, we adopt a representation for bent notes consisting
of four different labels. We also need to pre-process our data to obtain bend-less scores, and musical features
that could be used to train a machine learning model. Those considerations are presented hereafter.

\subsection{Labeling}

In the introduction, we presented the different types of bends that can
be encountered in guitar tablatures. Based on this taxonomy, we define 4 labels
that represent the motion and current state of the played string:
\begin{itemize}
    \item \nobend --- the string is not bent;
    \item \up --- the string is bent, causing the pitch to go up;
    \item \held --- the string was bent previously and is plucked again in that state. The pitch is
    constant, but is not the one expected from the note's string/fret position;
    \item \down --- the string was bent and is released, making the pitch go down accordingly.
\end{itemize}

We define those labels to circumvent the issue with \textit{up \& down} and \textit{complex bends}
that are not transcribed consistently. Labeling notes with the associated string movements therefore permits
representing bends accurately, without loss of generality.
%
Using these labels, all non-bent notes 
will be labeled $\varnothing$, \textit{basic upward bends} are labeled $\uparrow$,
\textit{held bends} correspond to $\rightarrow$, and \textit{reverse bends} are $\downarrow$.
To fit with this representation, \textit{up \& down bends} are split into two 
notes of equal duration, the first one being labeled
with $\uparrow$, and the second one with $\downarrow$.
An example of such labeling on an actual guitar track is shown on top of \autoref{fig:bendTypes}.

While bends can be of different amplitude, we do not include that information in our
labeling, as we do not aim at predicting it in this work 
but only focus on predicting when bends occur.
Similarly, we do not distinguish single notes from chords when predicting bends.
This means that when the system predicts the presence of a bend in a chord,
it does not specify on which string it occurs. The impact of this simplification 
is however limited, as bends rarely occur inside chords (12\% of bend events are found in chords in our corpus).

\subsection{Deriving a bend-less score simplification}
\label{ssec:removing_bends}
\begin{figure}
    \centering
    \includegraphics[width=\linewidth]{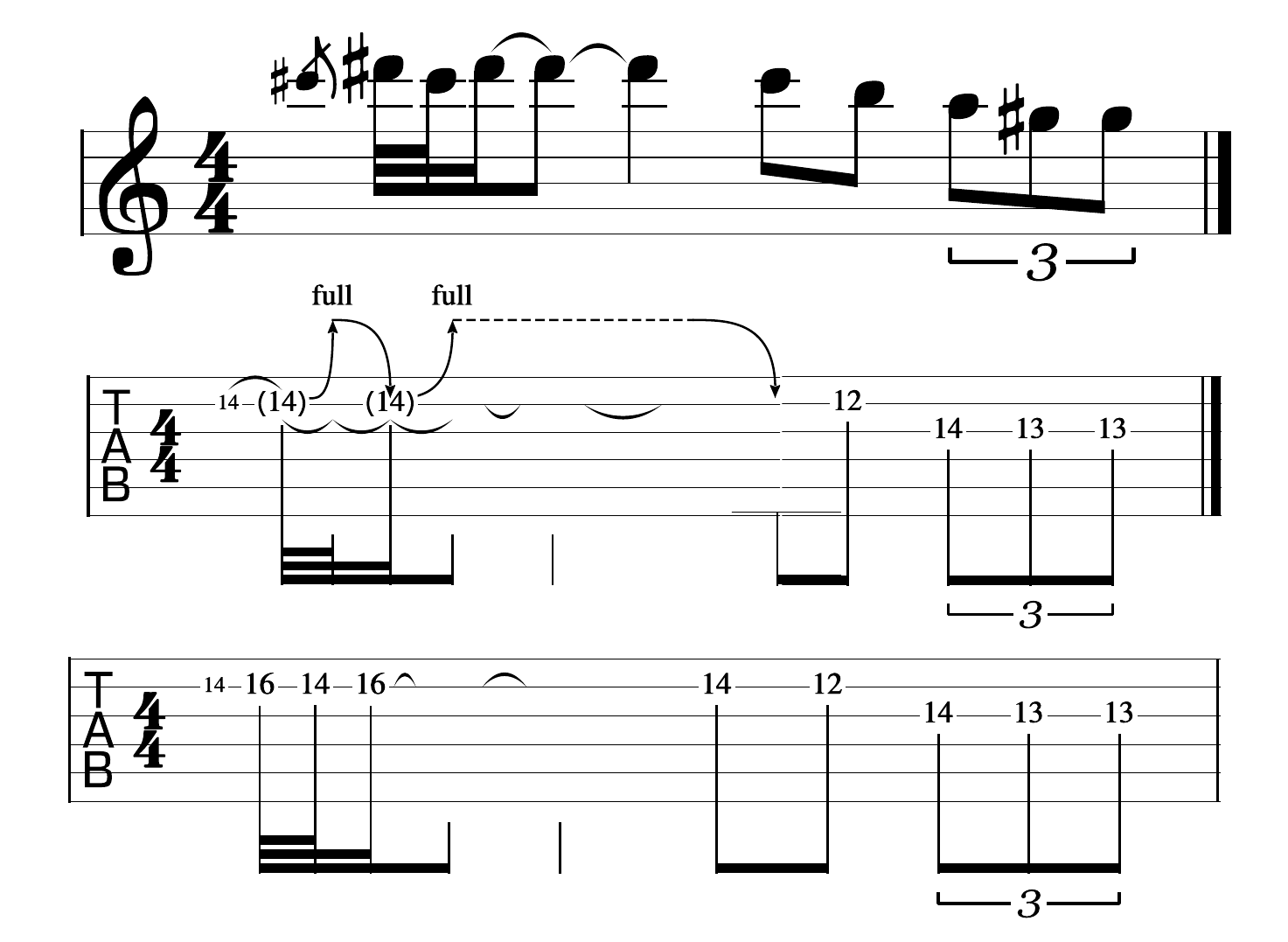}
    \caption{Excerpt of \emph{Watermelon in Easter Hay}, Frank Zappa, as transcribed \cite{zappa_frank_2017} by Steve Vai in standard notation (top).
    Below are two possible tablature representations of this excerpt with (middle), or without (bottom) bends.}
    \label{fig:zappa}
\end{figure}

Since our goal is to predict whether a note is played with a bend, 
we must start from a \emph{simplified} 
tablature that does not have any bend information. Removing 
bend annotations from a tablature is however not a straightforward
task since this technique affects the pitch of the performed note.
We design the following procedure when translating
bent notes from the fret/string space to the pitch space:

\begin{itemize}
    \item If a note is labeled by $\varnothing$, its pitch is directly obtained from the string/fret combination;
    \item Otherwise, the pitch is the one of the bend arrival note. In particular:
    \begin{itemize}
        \vspace{-1ex}
        \item if the label is \up or \held, the arrival pitch is the pitch of the string/fret combination,
        to which the bend amplitude is added;
        \item if the label is \down, the arrival pitch is the one corresponding to the string/fret position,
        because the string is released to its default state.
    \end{itemize}
\end{itemize}
This approach is the one chosen by Steve Vai when transcribing Frank Zappa's melodies
to standard notation, as we illustrate in \autoref{fig:zappa}.
The middle tablature shows how this excerpt
is actually played (based on a live performance) 
and the bottom tablature illustrates how the same excerpt might be played
without bends. 
While it is not an issue in this example, there is an uncertainty regarding where a bent note would be played on the fretboard,
without a bend (keeping its destination pitch but losing the technique).
A guitar player might indeed choose to play a note on a higher string, if remaining on the same
string called for an uncomfortably large hand span. 
Because deciding arbitrarily of a hand position could
introduce bias into our model, we choose not to include any
string/fret information concerning the current note in the proposed features for our
classification task, as explained hereafter.

\subsection{Selected Features}

To predict whether a note is bent, we propose an intermediate representation as a set of high-level features, presented in \autoref{tab:features}.
Some descriptors focus on the event under scrutiny, while others provide short-term context information, both from the past and the future.
Part of the features are derived from standard staff notation and convey \emph{temporal} and pitch information, while others are related
to \emph{position} and the tablature space. 
If the studied event is a chord, the pitch, fret, and string values are averaged over all its constituting notes. 
While the average might seem an overly simple statistic, experiments with other functionals such as \emph{min, max} or \emph{standard deviation}
did not improve the results. 
We have discarded any open strings for the fretboard features so that the average fret and string
represents the actual position of the fretting hand.
Apart from absolute/relative duration values, temporal features include the \textit{beat strength}, which is a value between 0 and 1 suggesting how \textit{strong}
the beat of a note is. We obtain this value from the default implementation of the \texttt{music21} library \cite{cuthbert_music21_2010}. 
These beat strength values have been
designed for Western classical music, 
and therefore may be debatable for pop
and rock music. However, they are mostly used here to represent onset times independently of the time signature, while grouping
notes that share rhythmic properties.

\begin{table}
\centering
\small
\begin{tblr}{
  cell{1}{1} = {r=5}{},
  cell{6}{1} = {r=5}{},
  cell{11}{1} = {r=4}{},
  hline{6,11} = {-}{},
  stretch=0
}
Temporal & Duration                      \\
                                       & Beat Strength                 \\
                                       & Longer than previous          \\
                                       & Shorter than previous         \\
                                       & Same duration as previous     \\
Pitch    & Number of notes               \\
                                       & Pitch$^{(j)}$                 \\
                                       & Pitch jump$^{(\mathrm{n}\pm k)}$       \\
                                       & Accidentals                   \\
                                       & Pitch-class w.r.t scale root       \\
Position & Fret$^{(\mathrm{n}\pm k)}$             \\
                                       & String$^{(\mathrm{n}\pm k)}$           \\
                                       & Fret jump~$^{(\mathrm{n}\pm 2)}$       \\
                                       & String jump~$^{(\mathrm{n}\pm 2)}$     
\end{tblr}
    \caption{List of high-level features extracted from the note events. Let $\mathrm{n}$ denote the current note index, an exponent on a feature tells on which neighbors it is computed. $k\in \{1,2\}$ because we discard any positional information related to the current note, and $j\in\llbracket\mathrm{n}-2, \mathrm{n}+2\rrbracket$. Other features are only computed on $\mathrm{n}$.}
    \label{tab:features}
\end{table}

In addition to the features on the current note, we extract a context
of two past and two future note events, as preliminary experiments did not show any benefits of longer contexts.
Additional boolean features are provided to recall 
if a neighboring note event is missing, when a note is preceded or followed by
a whole rest for instance. When a note is missing, all corresponding features are set to 0.
From this context, we compute the pitch jump between neighboring 
notes as well as the string and fret jumps when they are defined, \textit{i.e.,}
not with respect to the current note -- because we do not know where the guitarist would play the note
if they were to bend it. We expect these features to help our algorithm 
derive the hand position on the fretboard, which would be useful since bends are
more likely to occur on certain spots of the fretboard, as will be shown \autoref{ssec:stats}. 
Furthermore, we add information about the key signature through the number of accidentals (positive
for sharps, negative for flats). From those accidentals, we derive the root note of the corresponding pentatonic 
minor scale (that scale encompassing much of guitar popular music\cite{temperley_musical_2018}) and store the position of each note on this scale. For example,
one sharp would make the root E, and an A would be numbered 5 since it is 5 semitones above E.


\section{Dataset}
\label{sec:dataset}

\subsection{Guitar Tablature Corpus}

Our experiments are performed on the proprietary corpus \emph{MySongBook}
composed of 2247 guitar tablatures accurately transcribed by professional musicians
in the \texttt{.gp} GuitarPro format. A subset of 932 tracks estimated as \emph{lead guitar} -- totaling more than \num{130000} notes -- was 
extracted by applying the classification technique from \cite{regnier_identification_2021}. Our experiments focus
on lead guitar parts, as they were felt to feature heavier use of playing techniques. In contrast with the whole corpus, which includes \SI{2.5}{\%} of bent notes, our lead guitar sub-corpus indeed contains 10\% of bent notes, slightly mitigating the observed class imbalance.

Our work is implemented in \emph{Python} and uses \texttt{music21} \cite{cuthbert_music21_2010} and
\texttt{scikit-learn} \cite{pedregosa_scikit-learn_2011} libraries. To foster reproducibility,
all our code is made publicly available (parsing of \texttt{.gp} files, extraction of features, training, and evaluation of bend
classification models).
We also release the complete set of features extracted on each note of our corpus, plus corresponding labels at:\\ {\url{http://algomus.fr/code/}}. 


\subsection{Statistical study}
\label{ssec:stats}

\begin{table}
	\centering
	\begin{tabular}{cccc|c}
		  \nobend & \up & \held & \down & Total \\
		\hline
		  \num{123231} & \num{9627} & \num{1270} & \num{3314} & \num{137442}
	\end{tabular}
	\caption{Number of notes per label in our dataset.}
	\label{tab:bendStats}
\end{table}

\begin{figure}
    \centering
    \includegraphics[width=\linewidth]{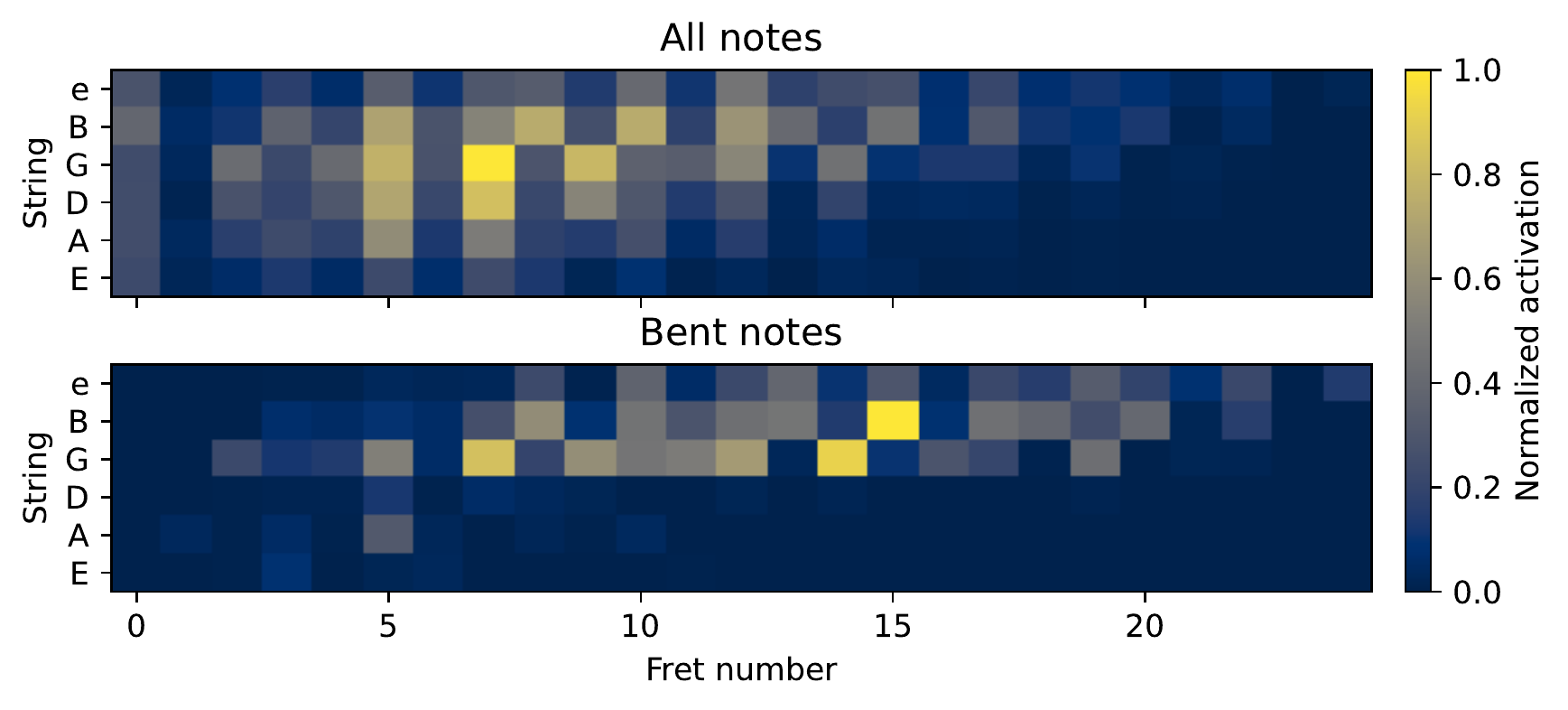}

    \caption{Normalized Heatmaps of all notes (Top) and bent notes (Bottom). The letters refer to the open string pitch in standard tuning with \emph{e} being the high E string.}
    \label{fig:heatmap}
\end{figure}

\begin{figure*}
    \centering
    \begin{subfigure}{.24\linewidth}
    \includegraphics[width=\linewidth]{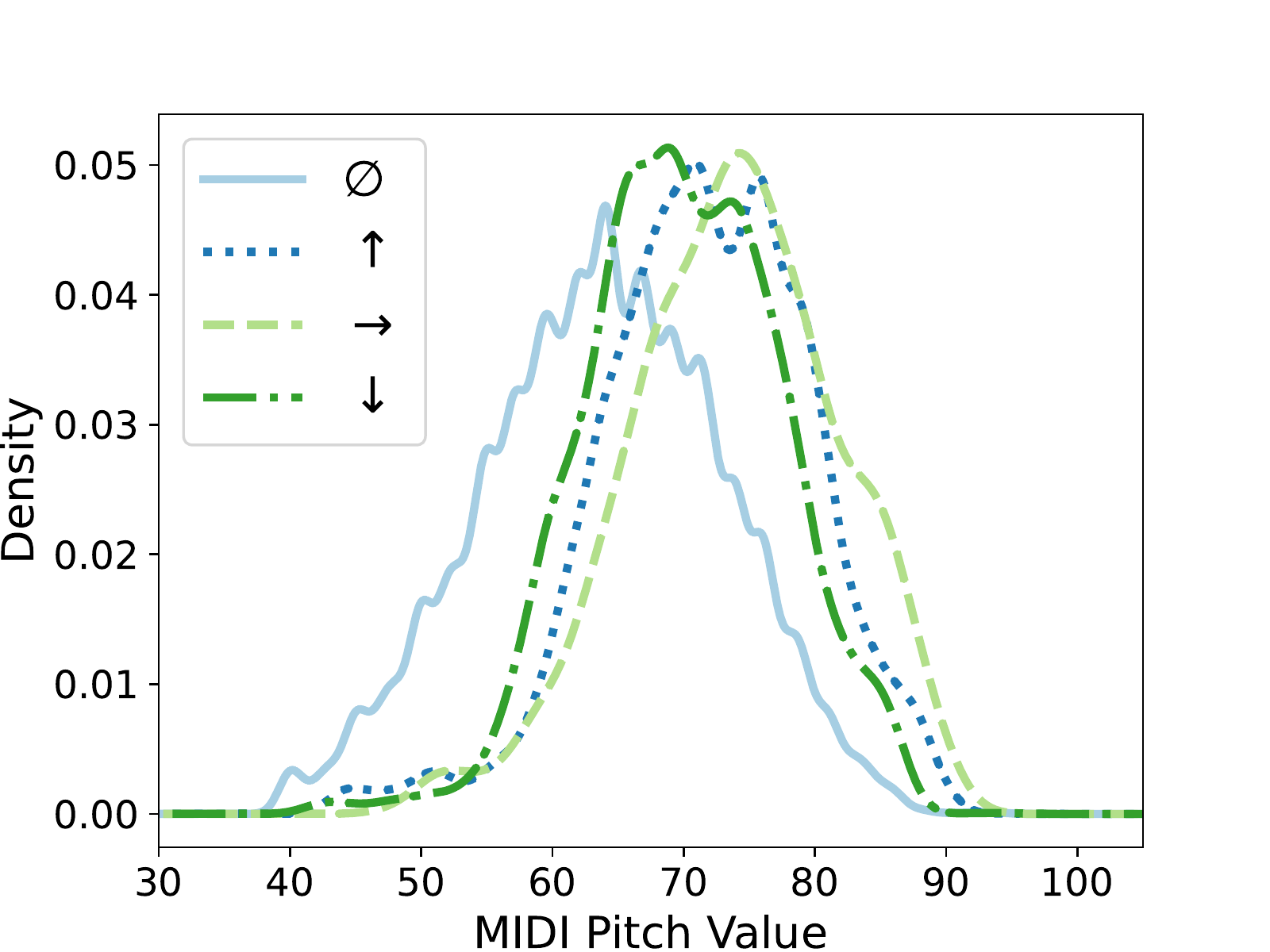}
    \caption{Pitch}
    \label{fig:meanPitchDensity}
    \end{subfigure}
    \hfill
    \begin{subfigure}{.24\linewidth}
    \includegraphics[width=\linewidth]{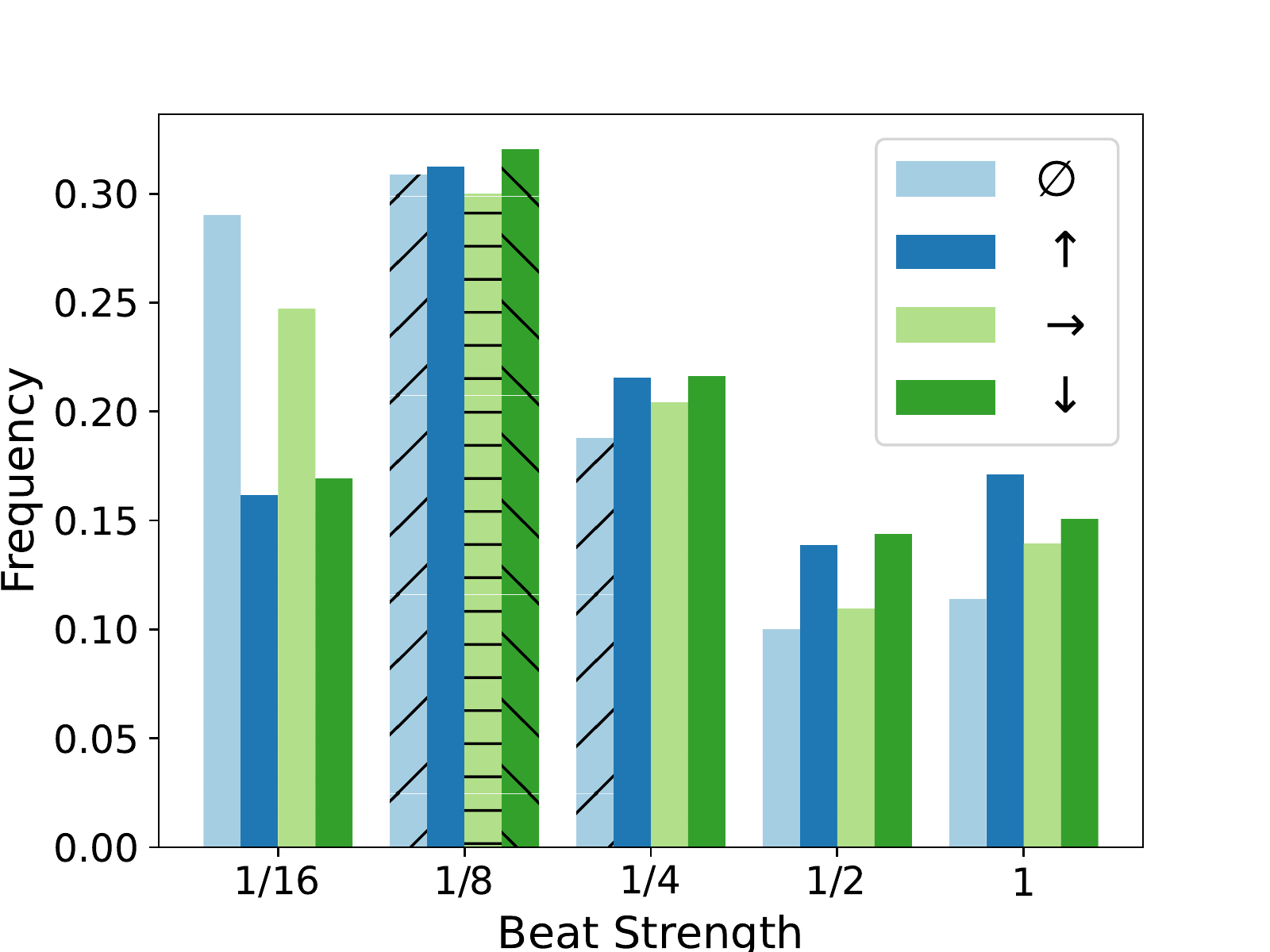}
    \caption{Beat Strength}
    \label{fig:beat_strength}
    \end{subfigure}
    \hfill
    \begin{subfigure}{.24\linewidth}
    \includegraphics[width=\linewidth]{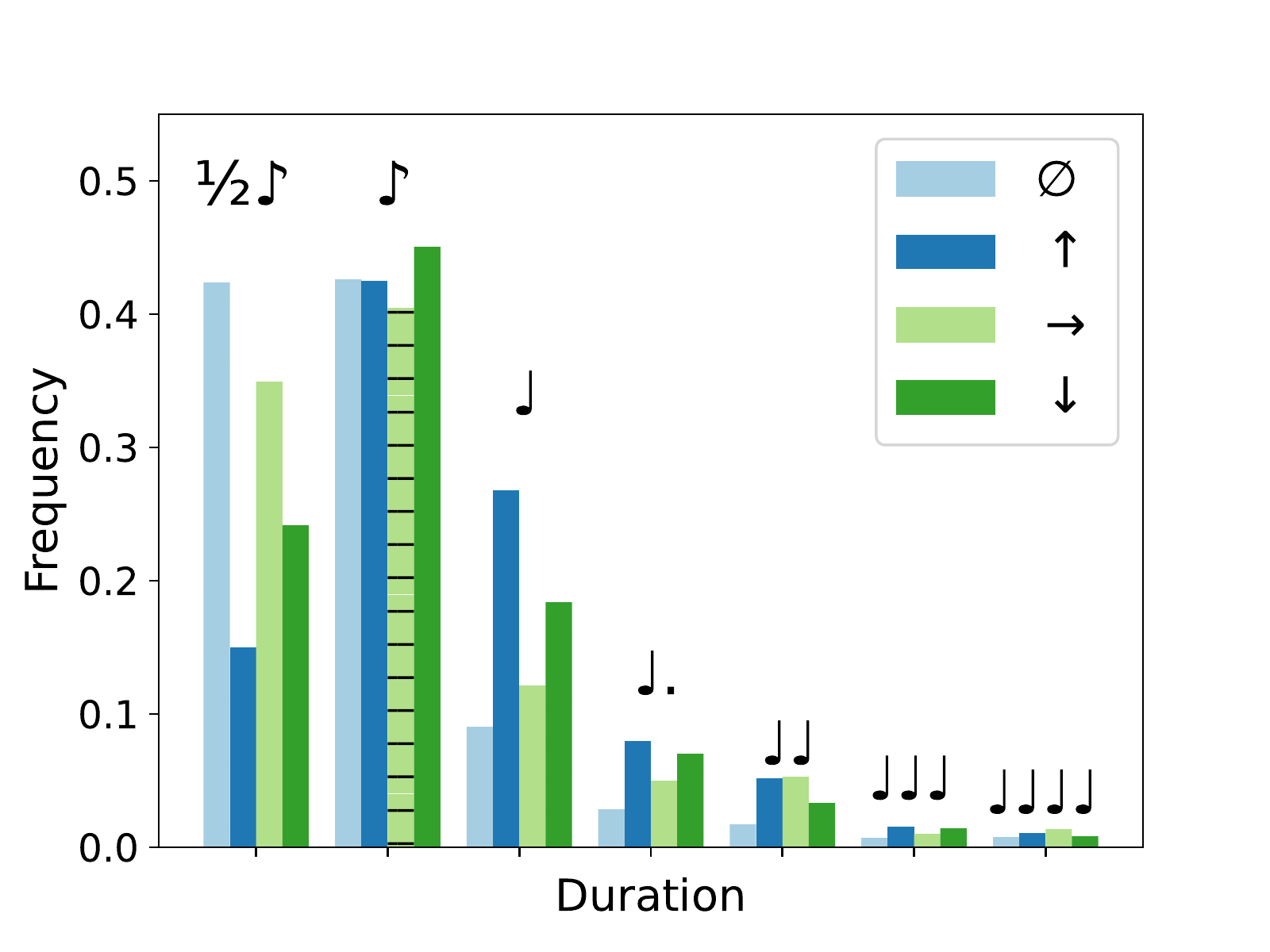}
    \caption{Duration}
    \label{fig:duration_hist}
    \end{subfigure}
    \hfill
    \begin{subfigure}{.24\linewidth}
    \includegraphics[width=\linewidth]{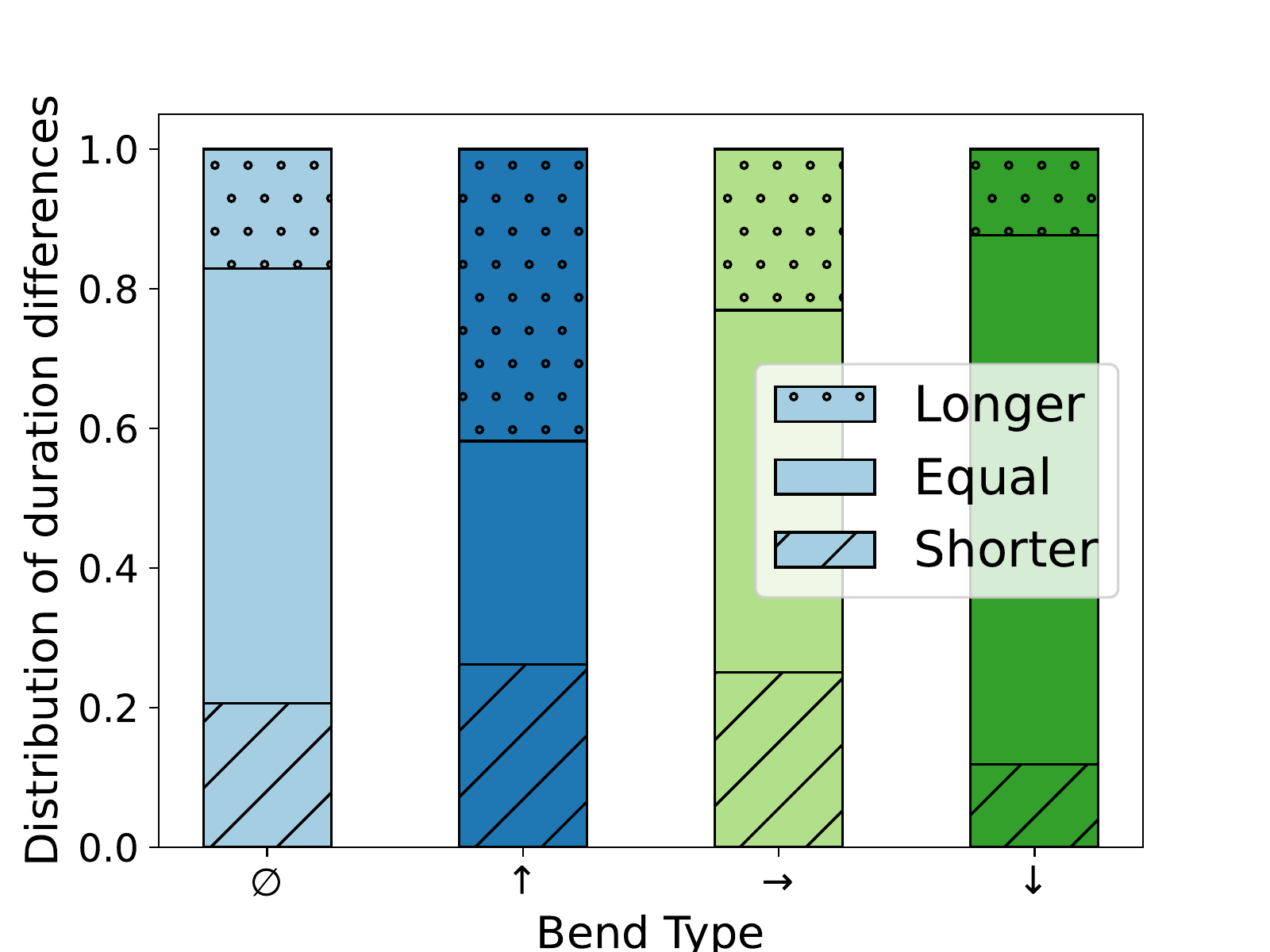}
    \caption{Duration diff. with previous}
    \label{fig:duration_diff}
    \end{subfigure}
    \caption{Distribution of four of the extracted features, normalized per bend class.}
    \label{fig:stats}
\end{figure*}

\autoref{tab:bendStats} reports the distributions of bend labels in our corpus. 
The distribution of bent notes on the fretboard, compared to all notes, is shown in \autoref{fig:heatmap}.
We observe that most bends occur on the top 3 strings in the middle area of the fretboard. 
This observation differs from notes in general that are played on all strings, and especially on the two middle ones and around the 7\textsuperscript{th} fret.
While it is possible that the obtained heatmaps are biased by an over-representation of certain
key signatures in the dataset -- 43\% of the tracks are in G major/E minor or C major/A minor -- 
this bias should affect both heatmaps equally, so their mutual comparison is still possible.
Because bent notes are found on both higher strings and higher frets than all notes,
their pitch is similarly higher on average, as it can be observed in \autoref{fig:meanPitchDensity}.

The distribution of \emph{beat strength} values is shown in \autoref{fig:beat_strength}. 
Because most beats and sub-beats in a measure have a beat strength of \num{0.25} or below,
no label is mostly played on strong beats.
An interesting result is that $\uparrow$ and $\downarrow$ labels appear more often on stronger beats than $\varnothing$ and $\rightarrow$. This apparent correlation of \up and \down labels with the meter might suggest a link between note expressiveness and accentuation in performance, which would need to be investigated further. In contrast, the  
$\rightarrow$ label is most often encountered on \textit{weaker} beats. This observation 
can be linked to the fact that this technique is often used as a quick repetition of the previous note 
and will thus be played on the next offbeat, like in \autoref{fig:bendTypes}. 

The comparison of the duration of notes with or without bends (\autoref{fig:duration_hist}) confirms that $\uparrow$ and $\downarrow$ labels share some essential properties.
Both labels have a proportionally higher tendency to be found on notes with longer duration,
even though eighth note is the most common duration for all classes.
This figure also confirms that \nobend and \held classes share some context properties. 
\autoref{fig:duration_diff} shows a strong tendency of \up labels to appear on notes with longer duration than their predecessor. 
This further supports the hypothesis that bends could be used to emphasize
significant notes in a lead guitar part. That result could also be related to the 
substantial physical effort required to bend a string on short duration notes.
The accompanying code provides interactive computation of the distribution of the other features.

\section{Classification results}
\label{sec:results}

A decision tree \cite{breiman_classification_2017} was trained to predict the bend label of a note from its feature representation. We choose this high-level approach to facilitate the interpretation
of the results as well as the analysis of the contribution of the features. 
In addition to the elaboration of a predictive model, conducting our experiments in an explainable AI framework allows us to improve our understanding of the use of bends in this repertoire. 
We hope that the use of light models enabled by highly expressive musical representations will also 
contribute to promoting low energy consumption approaches in machine learning for MIR.

\subsection{Model performance}

Our classifier is trained on 75\% of the dataset, and evaluated on the remaining 25\%.
To avoid some leakage from the training set to the test set, we ensure the split does not separate notes from a same track. 
We also ensure that the class imbalance is similar in both sets.
Duplicate feature vectors are removed track-wise
to avoid overfitting due to repeated riffs/patterns. But, we acknowledge the fact that identical feature vectors can be found in different tracks and 
thus keep duplicates when found in different files.

\begin{figure}
    \centering
    \includegraphics[width=.8\linewidth]{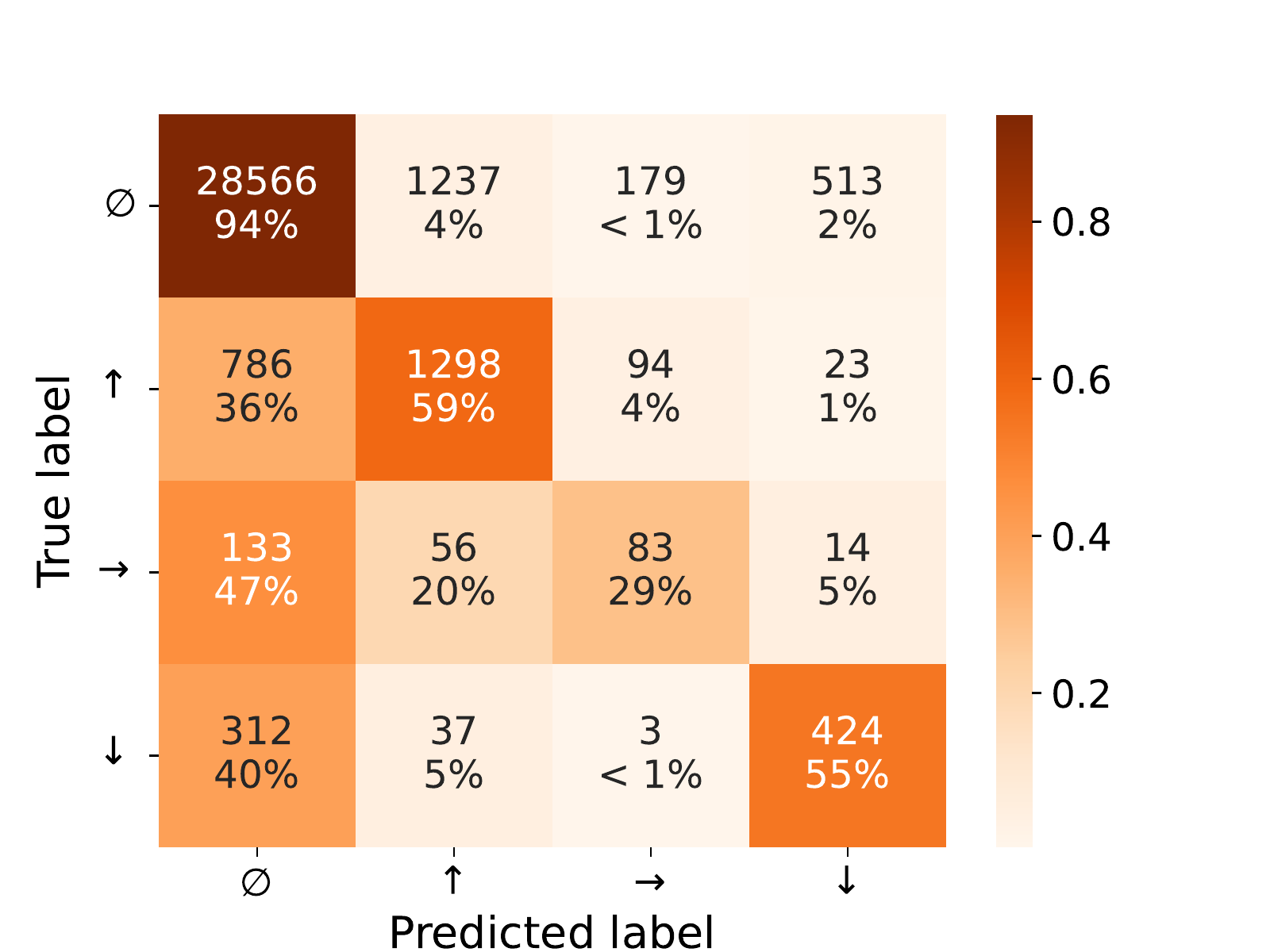}
    \caption{Confusion matrix obtained for classifying each note event to one of the bend class.
            This matrix was obtained on a split with average performance.}
    \label{fig:4class_classif}
\end{figure}

The confusion matrix of \autoref{fig:4class_classif} shows the results of the multi-class classification on the joint prediction of all labels. Because the dataset is highly unbalanced, our model is naturally biased towards the \nobend label.
However, it successfully identifies more than half of the \up and \down labels. 
Samples labeled as \held are often misidentified as \up but this result still shows, presumably,
that the model captures the difference between \nobend and \held labels.
We tried applying SMOTE oversampling \cite{chawla_smote_2002} to the training data
and observed that it doubles the number of correctly
identified \held notes and increases the ratio of well-classified \down notes by approximately 10\%.
Nevertheless, \up notes \emph{True Positives} (TP) ratio is about the same while
the quantity of \up notes misidentified as \held or \down increases.
Similarly, TP ratio of \nobend notes drops by 5 p.p., so 2000 more notes are
wrongly predicted as bent. Because we observed that bent notes are sparse in guitar tracks, 
we consider that \emph{precision} is more important than \emph{recall} and do not use any oversampling
for the rest of our analysis.

\subsection{Feature importance}

\begin{figure}
    \centering
    \includegraphics[width=.75\linewidth]{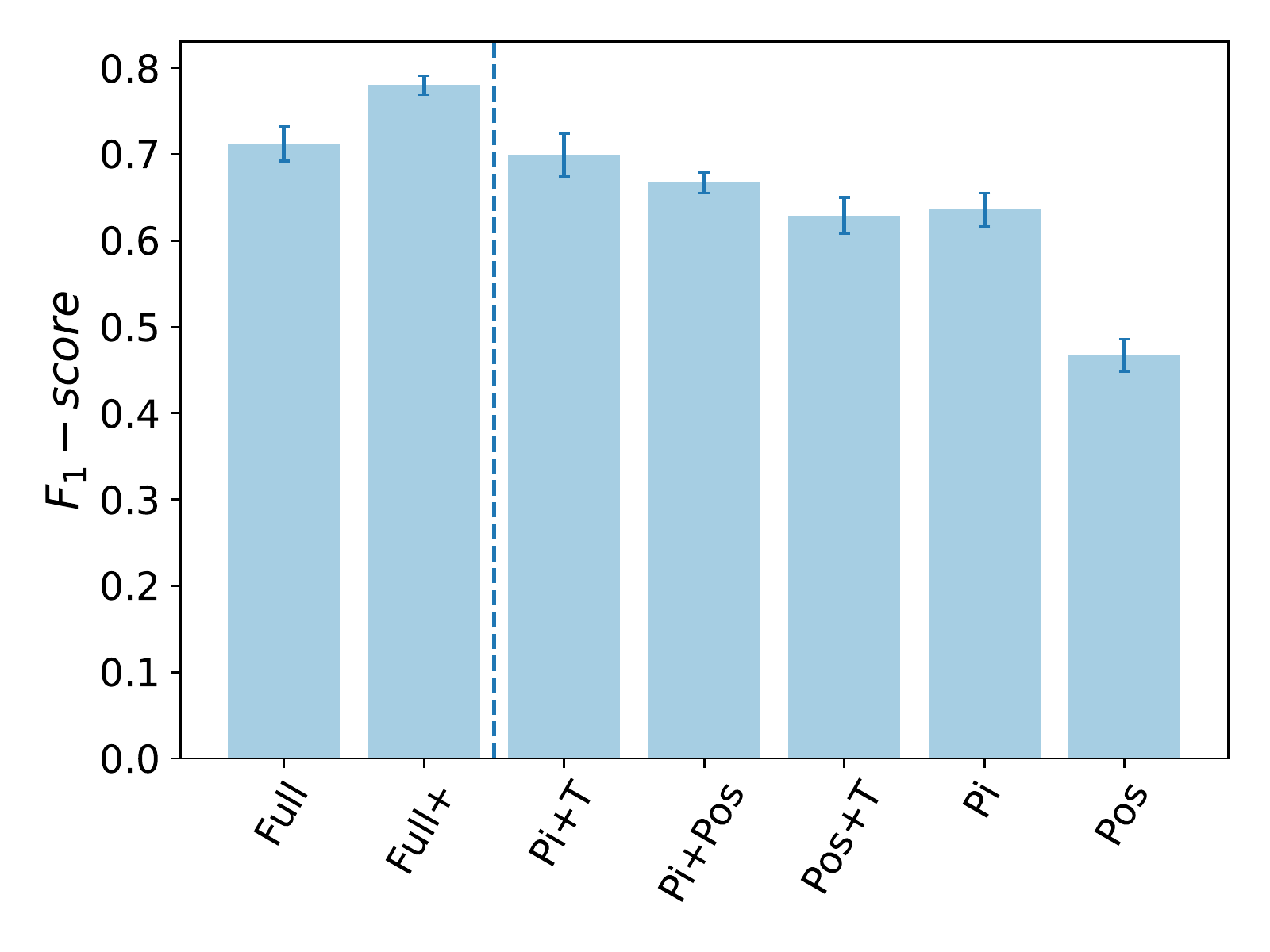}
    \caption{Average F$_1$-scores from 4 different train/test splits for the binary classification
    task. The leftmost part 
    shows the performance of the decision tree trained on all features, with (Full+) or without (Full) SMOTE oversampling.
    The rightmost part corresponds to decision trees trained with a reduced set of features. \emph{T} stands for temporal
    , \emph{Pi} for pitch and \emph{Pos} for position features.}
    \label{fig:fscores}
\end{figure}

To assess the contribution of each feature, we conduct  an \emph{all bend} binary classification experiment
where $\uparrow, \rightarrow, \downarrow$ are merged into a single class, versus the \nobend class. 
\autoref{tab:features_importance} shows the importance of the eight most contributing features, computed using the random feature permutation technique introduced in \cite{breiman_random_2001} and monitoring its impact on the F$_1$ score of our model.
Temporal and pitch features appear to have a higher impact on classification
than position-related features, an observation confirmed by training the binary classifier on selected subsets of features. 
The results in \autoref{fig:fscores} confirm the dominant influence of pitch features. 
However, adding gesture and temporal information noticeably improve the results.
This result suggests that, while fret context contributes to induce bent notes, a large part of the prediction can be done from the strict musical content as notated in musical scores.

\begin{table}[]
    \centering
    \begin{tabular}{l|c}
     \multicolumn{1}{c|}{Feature}  &  Importance \\
     \hline
     Pitch & \num{0.20} \\
     Pitch jump$^{(\mathrm{n}+1)}$ & \num{0.17} \\
     Pitch jump$^{(\mathrm{n}-1)}$ & \num{0.16} \\
     Duration & \num{0.14} \\
     Same dur. as previous & \num{0.07}\\
     Fret jump$^{(\mathrm{n}-2)}$ & \num{0.07}\\ 
     String$^{(\mathrm{n}+1)}$ & \num{0.05}\\
     Pitch$^{(\mathrm{n}+1)}$ & \num{0.05}\\ 
    \end{tabular}
    \caption{Feature importance of the 8 most significant features for the decision tree. Standard deviation
    of any feature importance is never above \num{0.005}.}
    \label{tab:features_importance}
\end{table}

\begin{figure*}[th]
    \centering
    \begin{subfigure}{\linewidth}
    \centering
    \includegraphics[width=.75\textwidth]{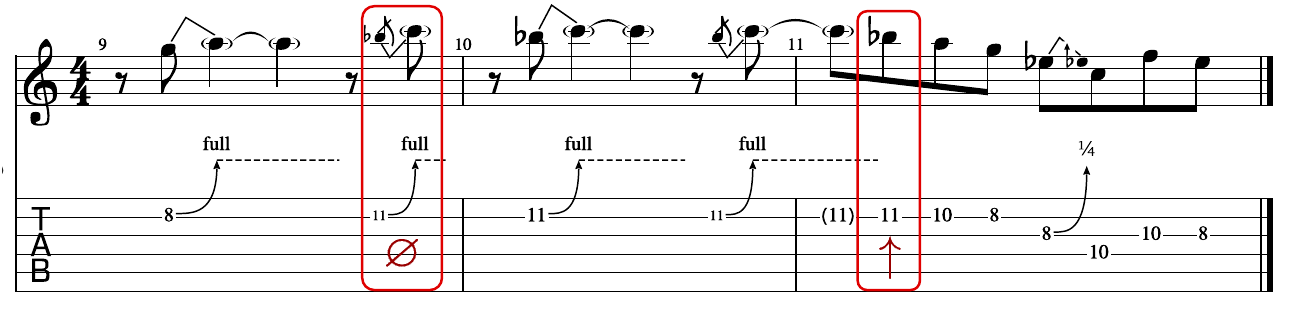}
    \caption{Excerpt from \textit{Highway Star}, Deep Purple.}\label{fig:highway}
    \end{subfigure}

    \begin{subfigure}{\linewidth}
    \centering
    \includegraphics[width=.80\textwidth]{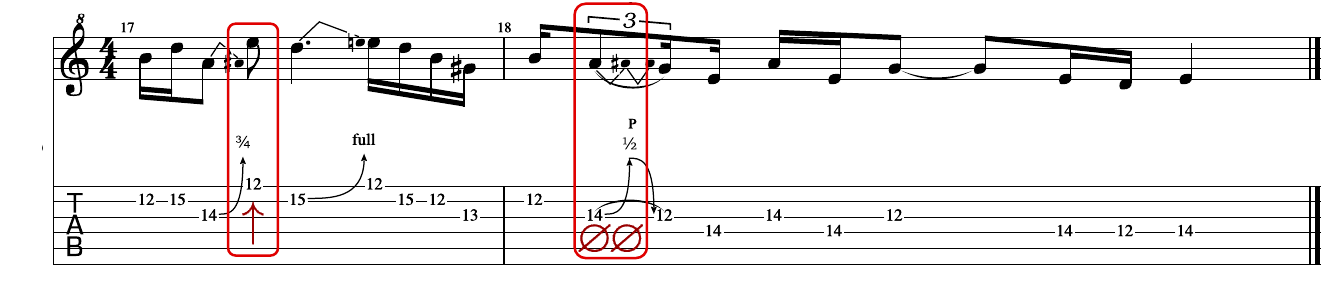}
    \caption{Excerpt from \textit{Jailbreak}, AC/DC.}\label{fig:jailbreak}
    \end{subfigure}
    \caption{Examples of predictions obtained with our Full Tree model on two different excerpts.
    Labels shown represent
    the predicted label for the current note. Only wrong predictions are shown for clarity. All other notes are labeled correctly.}
    \label{fig:pred_examples}
\end{figure*}

\section{Prediction analysis}
\label{sec:discussion}

In addition to the quantitative results presented in the last section, we present
a qualitative analysis of selected predictions.
\autoref{fig:highway} shows one bent note wrongly
identified as \nobend and, conversely, one
non-bent note identified as \up. 
Following the decision path provided by the decision tree, 
we can gain some insight on what feature differences
have caused those wrong predictions. 
Both notes actually have more than half their decision path in common and split on their 
Pitch jump$^{(n+2)}$ value, suggesting that the first discrepancy was due to future context.
In particular, the second false prediction did not use any features related to past context. 
This might also explain this error because the pitch could be obtained by bending on the 10\textsuperscript{th} fret by one semitone
-- an information that could be derived from future context -- 
but continuity with the previous notes called for playing the note without bend on the 11\textsuperscript{th} fret -- an information that should have
been derived from past context.
Another observation is that, in spite of similar context, the second bent note was misidentified whereas the fourth bent note was not.
While those two notes look very similar at first glance, the latter has a longer duration because it is tied 
to the following eighth note, which illustrates the importance of the \emph{duration} feature
An analysis of the decision paths indeed shows a divergence from the second decision rule, based on that feature. This highlights the presumably strong influence of rhythm in the classification of the first four bent notes, which bypasses pitch features.

\autoref{fig:jailbreak} also shows a regular note wrongly tagged with a \up label. 
The decision path for this prediction does not consider any feature related to the next note. It does however
use many features concerning the second next note, which was correctly classified
as \nobend, most likely because of its lower duration. The lack of information about the current note's position
was probably critical in that case.
The second error on that tablature is an \emph{up \& down bend} that was not identified, probably because of the low duration of the involved notes. Nevertheless,
this example suggests that our method to obtain a bend-less transcription from an up \& down bend might be detrimental
to the algorithm performance. Indeed, our procedure has an impact on \emph{duration} and \emph{pitch jump$^{(\mathrm{n}\pm1)}$}
which are among the most useful features to our algorithm. 
We observe however that our algorithm predicted correctly six bend labels in the selected examples with a limited
amount of false positives. These encouraging results suggest that our method could be used as a suggestion tool for the idiomatic use of bends.

\section{Conclusion}

In this paper, we proposed a model of guitar bends and discussed how these expressive playing techniques relate to both tablature and score content.
Introducing a set of high-level features, we showed that a decision tree can successfully predict bend occurrences with
satisfactory precision, in spite of the difficulty of the task due to the low proportion of bent notes in guitar music.
In particular, the low performance on predicting \held labels suggests that our modeling choices
could be improved and that held bends might not be considered as an expressiveness technique but rather 
another way of playing regular notes. 
An advantage of our approach is the use of a lightweight and explainable algorithm, facilitating its use in an assisted-composition context.
In future work, this approach could be extended to other guitar playing techniques,
and might benefit from adding more context information like the chord being played over a bar, using
rhythm guitar parts aligned with lead guitar. 
Because bends are arguably more easily performed with the ring finger and little finger than other fingers of the fretting-hand, combining our work
with finger prediction technique~\cite{hori_three-level_2021} might also improve prediction performance. 
Finally, our modeling strategy could also be used to study the playing style of specific guitarists,
and evaluate the potential of bends for automatic guitarist identification. Indeed, our approach supposes
that bends can be explained by general musical features regardless of the artist.
This is debatable since famous players can be identified by their solos (without considering audio
nor any playing technique) \cite{das_analyzing_2018}. It would be interesting to 
see if bends are artist-dependent and, if so, to develop a model that predicts
bends in the style of a specific guitarist. 




\textbf{Acknowledgements.} This work is made with the support of the French National
Research Agency, in the framework of the project TABASCO (ANR-22-CE38-0001).
The authors would like to thank Arobas Music for providing the dataset and 
colleagues from the Algomus team for their thorough proofreading and insightful comments.

\bibliography{TABASCO}

\begin{thebibliography}{10}
\providecommand{\url}[1]{#1}
\csname url@samestyle\endcsname
\providecommand{\newblock}{\relax}
\providecommand{\bibinfo}[2]{#2}
\providecommand{\BIBentrySTDinterwordspacing}{\spaceskip=0pt\relax}
\providecommand{\BIBentryALTinterwordstretchfactor}{4}
\providecommand{\BIBentryALTinterwordspacing}{\spaceskip=\fontdimen2\font plus
\BIBentryALTinterwordstretchfactor\fontdimen3\font minus
  \fontdimen4\font\relax}
\providecommand{\BIBforeignlanguage}[2]{{%
\expandafter\ifx\csname l@#1\endcsname\relax
\typeout{** WARNING: IEEEtran.bst: No hyphenation pattern has been}%
\typeout{** loaded for the language `#1'. Using the pattern for}%
\typeout{** the default language instead.}%
\else
\language=\csname l@#1\endcsname
\fi
#2}}
\providecommand{\BIBdecl}{\relax}
\BIBdecl

\bibitem{grimes_string_2014}
D.~R. Grimes, ``\BIBforeignlanguage{en}{String {Theory} - {The} {Physics} of
  {String}-{Bending} and {Other} {Electric} {Guitar} {Techniques}},''
  \emph{\BIBforeignlanguage{en}{Public Library of Science One}}, vol.~9, no.~7,
  Jul. 2014.

\bibitem{gomez_modern_2016}
P.~J. Gomez, ``\BIBforeignlanguage{en}{Modern {Guitar} {Techniques}; a view of
  {History}, {Convergence} of {Musical} {Traditions} and {Contemporary} {Works}
  ({A} guide for composers and guitarists)},'' Ph.D. dissertation, UC San
  Diego, 2016.

\bibitem{xi_guitarset_2018}
Q.~Xi, R.~M. Bittner, J.~Pauwels, X.~Ye, and J.~P. Bello,
  ``\BIBforeignlanguage{en}{Guitarset: {A} {Dataset} for {Guitar}
  {Transcription}},'' in \emph{\BIBforeignlanguage{en}{Proc. of the 19th
  {International} {Society} for {Music} {Information} {Retrieval}
  {Conference}}}, Paris, France, 2018.

\bibitem{wiggins_guitar_2019}
A.~Wiggins and Y.~Kim, ``\BIBforeignlanguage{en}{Guitar {Tablature}
  {Estimation} with a {Convolutional} {Neural} {Network}},'' in
  \emph{\BIBforeignlanguage{en}{Proc. of the 20th {International} {Society} for
  {Music} {Information} {Retrieval} {Conference}}}, Delft, The Netherlands,
  2019.

\bibitem{reboursiere_left_2012}
L.~Reboursière, S.~Dupont, O.~Lähdeoja, C.~Picard-Limpens, T.~Drugman, and
  N.~Riche, ``\BIBforeignlanguage{en}{Left and right-hand guitar playing
  techniques detection},'' \emph{\BIBforeignlanguage{en}{NIME}}, 2012.

\bibitem{chen_playing_2018}
S.-H. Chen, Y.-S. Lee, M.-C. Hsieh, and J.-C. Wang, ``Playing {Technique}
  {Classification} {Based} on {Deep} {Collaborative} {Learning} of
  {Variational} {Auto}-{Encoder} and {Gaussian} {Process},'' in \emph{2018
  {IEEE} {International} {Conference} on {Multimedia} and {Expo} ({ICME})},
  Jul. 2018, pp. 1--6.

\bibitem{yu-fen_huang_joint_2020}
{Yu-Fen Huang}, {Jeng-I Liang}, {I-Chieh Wei}, and {Li Su}, ``Joint analysis of
  mode and playing technique in {Guqin} performance with machine learning,'' in
  \emph{Proc. of the 21st {International} {Society} for {Music} {Information}
  {Retrieval} {Conference}}, Montréal, Canada, 2020.

\bibitem{sarmento_dadagp_2021}
P.~Sarmento, A.~Kumar, C.~J. Carr, Z.~Zukowski, M.~Barthet, and Y.-H. Yang,
  ``\BIBforeignlanguage{en}{{DadaGP}: {A} {Dataset} of {Tokenized} {GuitarPro}
  {Songs} for {Sequence} {Models}},'' in \emph{\BIBforeignlanguage{en}{Proc. of
  the 22nd {Int}. {Society} for {Music} {Information} {Retrieval} {Conf}.}},
  Online, 2021.

\bibitem{sarmento_gtr-ctrl_2023}
P.~Sarmento, A.~Kumar, Y.-H. Chen, C.~Carr, Z.~Zukowski, and M.~Barthet,
  ``\BIBforeignlanguage{en}{{GTR}-{CTRL}: {Instrument} and {Genre}
  {Conditioning} for {Guitar}-{Focused} {Music} {Generation}
  with {Transformers}},'' in \emph{\BIBforeignlanguage{en}{Artificial
  {Intelligence} in {Music}, {Sound}, {Art} and {Design}}}, ser. Lecture
  {Notes} in {Computer} {Science}, C.~Johnson, N.~Rodríguez-Fernández, and
  S.~M. Rebelo, Eds.\hskip 1em plus 0.5em minus 0.4em\relax Cham: Springer
  Nature Switzerland, 2023, pp. 260--275.

\bibitem{chen_automatic_2020}
Y.-H. Chen, Y.-H. Huang, W.-Y. Hsiao, and Y.-H. Yang,
  ``\BIBforeignlanguage{en}{Automatic {Composition} of {Guitar} {Tabs} by
  {Transformers} and {Groove} {Modeling}},'' in
  \emph{\BIBforeignlanguage{en}{Proc. of the 21st {International} {Society} for
  {Music} {Information} {Retrieval} {Conference}}}, Montréal, Canada, 2020.

\bibitem{mcvicar_autoleadguitar_2014}
M.~McVicar, S.~Fukayama, and M.~Goto,
  ``\BIBforeignlanguage{en}{{AutoLeadGuitar}: {Automatic} generation of guitar
  solo phrases in the tablature space},'' in \emph{\BIBforeignlanguage{en}{2014
  12th {International} {Conference} on {Signal} {Processing} ({ICSP})}}.\hskip
  1em plus 0.5em minus 0.4em\relax Hangzhou, Zhejiang, China: IEEE, Oct. 2014,
  pp. 599--604.

\bibitem{ching_learning_2021}
J.~Ching and Y.-H. Yang, ``Learning {To} {Generate} {Piano} {Music} {With}
  {Sustain} {Pedals},'' in \emph{Extended {Abstracts} for the {Late}-{Breaking}
  {Demo} {Session} of the 22nd {Int}. {Society} for {Music} {Information}
  {Retrieval} {Conf}.}, 2021.

\bibitem{sayegh_fingering_1989}
S.~I. Sayegh, ``Fingering for {String} {Instruments} with the {Optimum} {Path}
  {Paradigm},'' \emph{Computer Music Journal}, vol.~13, no.~3, pp. 76--84,
  1989.

\bibitem{hori_minimax_2016}
G.~Hori and S.~Sagayama, ``\BIBforeignlanguage{en}{Minimax {Viterbi} algorithm
  for {HMM}-based {Guitar} fingering decision},'' in
  \emph{\BIBforeignlanguage{en}{Proc. of the 17th {International} {Society} for
  {Music} {Information} {Retrieval} {Conference}}}, 2016.

\bibitem{cheung_semi-supervised_2021}
V.~K.~M. Cheung, H.-K. Kao, and L.~Su,
  ``\BIBforeignlanguage{en}{Semi-supervised violin fingering generation using
  variational autoencoders},'' in \emph{\BIBforeignlanguage{en}{Proc. of the
  22nd {International} {Society} for {Music} {Information} {Retrieval}
  {Conference}}}, Online, 2021.

\bibitem{hori_three-level_2021}
G.~Hori, ``\BIBforeignlanguage{en}{Three-{Level} {Model} for {Fingering}
  {Decision} of {String} {Instruments}},'' in
  \emph{\BIBforeignlanguage{en}{Proc. of the 15th {International} {Symposium}
  on {CMMR}}}, Online, 2021.

\bibitem{xie_symbolic_2020}
Y.~Xie and R.~Li, ``\BIBforeignlanguage{en}{Symbolic {Music} {Playing}
  {Techniques} {Generation} as a {Tagging} {Problem}},'' Oct. 2020, preprint.

\bibitem{zappa_frank_2017}
F.~Zappa and S.~Vai, \emph{\BIBforeignlanguage{eng}{The {Frank} {Zappa}
  {Guitar} {Book}}}.\hskip 1em plus 0.5em minus 0.4em\relax Hal Leonard
  Corporation, 2017.

\bibitem{cuthbert_music21_2010}
M.~S. Cuthbert and C.~Ariza, ``music21: {A} {Toolkit} for {Computer}-{Aided}
  {Musicology} and {Symbolic} {Music} {Data},'' in \emph{Proc. of the 11th
  {International} {Society} for {Music} {Information} {Retrieval}
  {Conference}}, 2010.

\bibitem{temperley_musical_2018}
D.~Temperley, \emph{\BIBforeignlanguage{en}{The {Musical} {Language} of
  {Rock}}}.\hskip 1em plus 0.5em minus 0.4em\relax Oxford University Press,
  Jan. 2018.

\bibitem{regnier_identification_2021}
D.~Régnier, N.~Martin, and L.~Bigo, ``\BIBforeignlanguage{en}{Identification
  of rhythm guitar sections in symbolic tablatures},'' in
  \emph{\BIBforeignlanguage{en}{Proc. of the 22nd {International} {Society} for
  {Music} {Information} {Retrieval} {Conference}}}, Online, 2021.

\bibitem{pedregosa_scikit-learn_2011}
F.~Pedregosa, G.~Varoquaux, A.~Gramfort, V.~Michel, B.~Thirion, O.~Grisel,
  M.~Blondel, P.~Prettenhofer, R.~Weiss, V.~Dubourg, J.~Vanderplas, A.~Passos,
  D.~Cournapeau, M.~Brucher, M.~Perrot, and E.~Duchesnay, ``Scikit-learn:
  {Machine} {Learning} in {Python},'' \emph{Journal of Machine Learning
  Research}, vol.~12, no.~85, pp. 2825--2830, 2011.

\bibitem{breiman_classification_2017}
L.~Breiman, \emph{Classification and regression trees}.\hskip 1em plus 0.5em
  minus 0.4em\relax Routledge, 2017.

\bibitem{chawla_smote_2002}
N.~V. Chawla, K.~W. Bowyer, L.~O. Hall, and W.~P. Kegelmeyer,
  ``\BIBforeignlanguage{en}{{SMOTE}: {Synthetic} {Minority} {Over}-sampling
  {Technique}},'' \emph{\BIBforeignlanguage{en}{Journal of Artificial
  Intelligence Research}}, vol.~16, pp. 321--357, Jun. 2002.

\bibitem{breiman_random_2001}
L.~Breiman, ``\BIBforeignlanguage{en}{Random {Forests}},''
  \emph{\BIBforeignlanguage{en}{Machine Learning}}, vol.~45, no.~1, pp. 5--32,
  Oct. 2001.

\bibitem{das_analyzing_2018}
O.~Das, B.~Kaneshiro, and T.~Collins, ``Analyzing and classifying guitarists
  from rock guitar solo tablature,'' in \emph{Proceedings of the {Sound} and
  {Music} {Computing} {Conference}}, Limassol, Chypre, 2018.

\end{thebibliography}

%
%
%
%
%

\end{document}